# Investigation of confined electron gas in a quantum well structure with photovoltaic application

*Partha Goswami*
*Deshbandhu college, University of Delhi, Kalkaji, New Delhi-110019,India*

*Garima Puneyani*
*Department of Electronic Sciences, University of Delhi South Campus, Delhi-110021,India*

*Avinashi Kapoor*
*Department of Electronic Sciences, University of Delhi South Campus,Delhi-110021,India*

**Abstract.** The growth-diection quantization of confined electron gas in a GaAs/AlGaAs based quantum well structure is obtained in the Kohn-Sham iterative computational scheme. The longitudinal conductance at low temperatures, in the presence of magnetic field applied in the same direction, is studied to confirm the presence of the bound-states obtained through this scheme. The sub-band absorption is calculated by a frequently used approximation modeling the absorption as a convolution between spectral functions. The absorption coefficient is used to obtain photovoltaic performance related parameters of a quantum well solar cell. When compared to a device without the quantum wells, it is found that the additional carrier generation in the wells due to extension of absorption edge into longer wavelengths brings about moderate enhancement in the values of these parameters.

# I. MAIN TEXT

The physics of optical transitions in semiconductor quantum well structures(SQWS) has attracted a great deal of interest since mid-eighties [1,2,3,4]. The reason is the confined electron gas(CEG) dispersion (or sub-bands), in the conduction band of these structures, are of practical interest. For example, it has been shown that the structures can lead to controlled photocurrent generation [5], electron inter sub-band transitions [6],etc. Motivated by these works, we investigated carrier generation process in SQWS through sub-band optical absorption. The aim is to gain access to carrier generation phenomenon in the quantum-confined structure, starting with a purely quantum mechanical description, and to calculate various photovoltaic performance related parameters (PPRP) of a generic quantum well solar cell(QWSC). For the latter, additionally, we solve the usual minority carrier transport equation for the emitter region. The sub-bands, resulting from the quantization of the one-electron states along z direction (in which the electron is trapped), are obtained in a straightforward manner through an iterative computational scheme (ICS), outlined below, based on time-independent Schrodinger equation for an electron, Poisson equation and the exchange-correlation potential $V_{xc}$ given by Gunnarson- Lundqvist form [7].

We consider a heterostructure of GaAs and $Al_x Ga_{1-x}$ As with the doping level x nearly 0.5 for the barrier material ($Al_x Ga_{1-x}$ As) so that the band gap mismatch is nearly 0.7 eV. For the case of a modulation-doped heterostructure one may assume the confining potential to be triangular. However, we consider a more frequently used confinement, viz. a smooth parabolic confinement, of the form $V(z) = [-V_0 + (1/2) m_z^* \omega_0^2 z^2]$ for $-d_z/2 < z < d_z/2$ (and zero elsewhere) with $d_z$ as the width of the well and $V_0$ to be of the order of the bulk band-gap mismatch of the well (GaAs) and the barrier material; $m_z^*$ ~0.07 $m_e$ is the effective mass for GaAs and $\omega_0 = (8V_0/ m_z^* d_z^2)^{1/2}$. As electrons are transferred from the barrier to the well region, the conduction and valence band edges are modified from the flat ones. This band-bending effect is included in the Hartree and the exchange-correlation potentials. Though not considered, an additional potential energy term due to a d.c electric field will be of advantage here, for then as the field increases confined carriers would tunnel out of the well lowering their potential energy. In our computational scheme, one starts with an initial guess for the electron density $n(z)$ examining the Poisson equations for the well and the barrier assuming impurity profile to be typically Gaussian shaped [8]. The Hartree and exchange-correlation potentials are then computed for this density. The Schrödinger equation is solved numerically to obtain $\varphi_j(z)$ and $\varepsilon_j$ where $\varphi_j(z)$ and $\varepsilon_j$ are the sub-band wave functions and energies respectively. A new density $n_{new}(z)$ is then computed using $\varphi_j(z)$ and compared to the previous one. If $\zeta$ ($= (\int dz\, |n_{new}(z) - n_{old}(z)|\,)/\int dz\, n_{old}(z)$) is larger than some specified tolerance(~ 0.01), we define a density $n_{NEW}(z) = [n_{new}(z)(1 - \eta) + n_{old}(z)\, \eta\,]$ where $0 < \eta < 1$. The density $n_{NEW}(z)$ is then used as input for the iteration procedure until $\zeta$ is smaller than the tolerance. It takes typically 35 iterations to achieve a value of $\zeta < 0.01$. The final $n(z)$ turns out to be $n_e = 2.6 \times 10^{16}$ m$^{-2}$. We remark that the scheme, which is reminiscent of the Density functional theory (DFT)[9,10], is good for the computation of ground state energy. Admittedly, an extension of DFT, viz. time-dependent DFT(TD-DFT) [11,12], is more suitable for the discussion of excited states. Since the formulation of the TD-DFT is found to be too complicated to solve the wave functions numerically, a considerably approximate formula called the TD-Kohn-Sham (TD-KS) equation has been applied for the numerical simulations [13] by previous workers. The difficulty in numerically solving the TD-KS equation is the treatment of the density-dependent Hamiltonian. The wave functions and the Hamiltonian should always be self- consistent with each other. These caveats are, however, not relevant for our scheme as its efficacy depends on how good the initial guess for the electron density is. We find that the estimation of ground state and excited state energies ($\varepsilon_j = -0.52$ eV, $-0.29$ eV, $-0.08$ eV ( for $d_z=10$nm, barrier width = 20 nm and $V_0= 0.62$ eV))are reasonable enough for the impurity concentration $N_d \sim 10^{22}$ m$^{-3}$. The energies ($-0.52$ eV, $-0.08$ eV) correspond to states with even parity whereas ($-0.29$ eV) to odd parity. Apart from these bound states, we also obtain a bunch of continuum levels (QCL)with positive energy. The optical absorption induced transitions to these ones from the bound states are relevant for the photovoltaic application. As the tunneling probability of an electron from quasi-bound levels to classically forbidden region being much higher, these transitions are more likely to bring about enhancement in the short-circuit current of a quantum well solar cell(QWSC).

In order to obtain signature of these bound states in the presence of strong magnetic field B ($\omega_0<\omega_c=eB/m^*$) applied in z-direction at low temperature, we have also studied Shubnikov de Haas(SdH) effect [14] in longitudinal conductance(G) calculated from one-particle energies $E_{N,j} = [(N+(1/2))\hbar\omega_c + \varepsilon_j]$ where N = 0,1,2,….We find that $G = \nu G_0$ where $\nu = (1/2)\int dE\,(-\partial/\partial E)f(E,T)\int_0^\pi d(qx_0)\sum_j \sum_{N=0}^\infty [(2N+1)/\{(N-\alpha_j)^2 +\gamma^2\}^2], \alpha_j=[\{(E+\varepsilon_j)/(\hbar\omega_c)\}-(1/2)-(U/\hbar\omega_c)\times\cos(qx_0)], \gamma=(\acute{\Gamma}/\hbar\omega_c)$, f(E,T) is the usual Fermi-Dirac distribution function and the Fermi energy $E_F=(\pi n_e \hbar^2/m^*)$ with $m^*$ being the transverse mass of electrons. The quantity $G_0 =(2e^2/h)$ is the Landauer conductance of a single, one-dimensional, ideal channel. The additional periodic potential energy $\{U\cos(qx)\}$ with $q=(2\pi/b)$ lifts the degeneracy of Landau levels and makes the energy become dependent on the position $x_0$ of the guiding centre. We have taken U~0.5 meV and the period b =150 nm. Here $\acute{\Gamma} =(\hbar e/2m^*\mu)$ corresponds to the level-broadening due to collisions and $\mu$ to the mobility. The details regarding the overlap of Landau levels $(N+(1/2))\hbar\omega_c$ and the levels $\varepsilon_j$ arising out of the spatial confinement at low but finite temperature, including the possibility of many-body states such as skyrmions for $\omega_0 <\omega_c$ in a semi-classical approximation where the spin-charge separation of the electron field is based on the composite-boson picture[15], will be reported elsewhere. Applying the Poisson sum formulae in the expression for G given above we have obtained the frequencies of SdH oscillations and obtained a confirmation of the energy values $\varepsilon_j$. For the values of B assumed here, the change in orbital energy $(\hbar eB/ m^*)$ is much larger than the Zeeman spin splitting since g-factor for GaAs is (−0.45) at low temperature (see Ref.[16]). In Figure 1 we plot the conductance G (at a given B) with the absolute value of the upper and lower limits of the energy integration above set at a value>>kT, in thermal Boltzmann units kT. Depending on the ratio of Fermi energy and temperature T, we find that the conductance duly undershoots its ideal, ballistic Landauer limit indicating the signature of the bound states; though the step-like structure ( in the Landauer limit) does not survive. We also notice that, for this three-bound-level(TBL) system, the inter sub-band transition frequencies are of order 100 THz while the transition from these state to continuum correspond to frequencies one order of magnitude greater.

We next outline the inter sub-band optical absorption in the super-lattice under consideration made up of such quantum wells (tailored as TBL systems). In a super-lattice (SL) or a multiple quantum well (MQW) system, there are two frequency regimes: $\hbar\omega < E_{ISBT}$, and $E_{ISBT} \leq \hbar\omega \leq E_g$, where $E_{ISBT}$ is inter sub-band transition energy and $E_g$ is the bulk band gap. The optical absorption of the nanostructure is calculated, through a convolution approach, for the low-frequency regime $E_{ISBT} \leq \hbar\omega \leq E_g$. We assume, for simplicity, that in the superlattice under consideration the neighboring potential wells are so far apart that the wavefunctions of the individual potential wells do not overlap. The transition rates corresponding to absorption and emission has been determined using the Fermi Golden Rule which states that the probability of a transition occurring per unit time, $\chi_{i \to f}$, is given by $\chi_{i \to f}(\omega) = (2\pi/\hbar) \left| \langle f | \gamma \hat{A} | i \rangle \right|^2 \partial(\varepsilon_f - (\varepsilon_i + \hbar\omega))$; the delta function expresses the fact that energy is conserved. The operator $\gamma \hat{A}$ is given by $(\hat{a}_{ij} + \hat{e}_{ji})$ where $\hat{a}_{ij} \equiv \sum_{\sigma, ij(i>j)} a_{ij}\, a_{i\sigma}^{\dagger} a_{j\sigma}$ (i,j=0,1,2,3), and $\hat{e}_{ji} \equiv \sum_{\sigma, ij(i<j)} e_{ij}\, a_{i\sigma}^{\dagger} a_{j\sigma}$ ; ($a_{ij}, e_{ij}$) are the inter sub-band optical transition matrix elements. Here the index 0 corresponds to the ground state, (1,2) to the excited states, and 3 to the continuum state. The creation operators $a_{j\sigma}^{\dagger}$ correspond to the $j^{th}$ sub-band and the index $\sigma$ to spin ; $\varepsilon_i$ (i = 0,1,2) are the energies corresponding to the three bound states and $\varepsilon_3$ to the continuum states. The correlation effect, if included by the term $(2U a_{i\uparrow}^{\dagger} a_{i\uparrow}\, a_{i\downarrow}^{\dagger} a_{i\downarrow})$ (= $U(a_{i\uparrow}^{\dagger} a_{i\uparrow} + a_{i\downarrow}^{\dagger} a_{i\downarrow}) + \Delta_1\, a_{i\downarrow}^{\dagger} a_{i\uparrow} + \Delta_2\, a_{i\uparrow}^{\dagger} a_{i\downarrow}$ where the last two terms involving the fields $\Delta_1 = -U\, a_{i\uparrow}^{\dagger} a_{i\downarrow}$ and $\Delta_2 = -Ua_{i\downarrow}^{\dagger}a_{i\uparrow}$ may be dealt with by the equation of motion method [17] in the Hubbard approximation framework), gives rise to absorption peak broadening[18,19]. The details concerning this issue will be taken up in a separate communication. The net transition rate is obtained by summing over both i and f and weighing the sum by a probability, involving eigen values of the density matrix of the system, such that the initial state of the system is |i⟩. This leads to the relation $P(-\omega) = \exp(-\beta\hbar\omega) P(\omega)$ where $P(-\omega)$ and $P(\omega)$, respectively, are the transition rates corresponding to emission and absorption. This is the equation of detailed balance. We see that the probability of emission is less than that for absorption at any finite temperature. This, in turn, indicates that the additional carrier generation and photocurrent expected in quantum wells due to extension of absorption edge into longer wavelengths would outweigh the accompanying drop in the terminal voltage arising from increased radiative recombination loss inside the wells. However, the investigation of non-radiative loss at the hetero-interfaces remains as the key issue to be clarified.

We have calculated the bulk absorption $\alpha_{bulk}(\varepsilon)$ through the convolution approach[20,21] between the electron and hole spectral functions, starting with the multi-band model of GaAs in Ref.[5], in the frequency regime $\hbar\omega \geq E_g$ where $E_g$ is the bulk band gap. We have considered an additional term $H_{pert} = \sum_{k,k',\sigma} \{\mu_k a_k \exp(-i\omega_k t) c^{\dagger}_{k'\sigma} v_{k'\sigma} + \mu_k a^{\dagger}_k \exp(i\omega_k t) v^{\dagger}_{k'\sigma} c_{k'\sigma}\}$ which describes the excitation (first term) and the annihilation (second term) of excitons in the system by photon absorption or photon emission, respectively. The operator $c^{\dagger}_{k'\sigma}$ ($v^{\dagger}_{k'\sigma}$) creates an electron of spin σ in the

conduction(valence) band. Here $a_k$ ($a^\dagger_k$) destroys (creates) a photon of the illuminating field with wave vector **k**, where the photon has the frequency $\omega_k = c|\mathbf{k}|$. The coupling of this field with matter is given by $\mu_k$ which we assume to be independent of **k.** The absorption peak ($\alpha_{bulk,max}(\varepsilon) \approx 1.75 \times 10^8$ m$^{-1}$ )occurs at photon energy $\varepsilon \approx 5.5$ eV. In Fig.2 we have shown the plot of $\alpha_{bulk}(\varepsilon)$-vs-$\varepsilon$ at 300K and the sixth-order curve fitting(depicted by dashed lines) of calculated values given by $(\alpha_{bulk}(\varepsilon)/(10^8 m^{-1})) = (0.0048\ \varepsilon^6 - 0.1141\ \varepsilon^5 + 1.0775\ \varepsilon^4 - 5.2091\ \varepsilon^3 + 13.649\ \varepsilon^2 - 17.971\ \varepsilon + 9.1652)$. The convolution approach enables us to avoid the evaluation of the two-particle (i.e. dipole-dipole) correlation functions leading to an important reduction of the numerical labor. It is straight-forward to see that in the convolution approach framework the sub-band absorption appears in the form $\alpha_{ISBT}(\varepsilon) \sim \sum_{ij(i>j)} A_{ij} \int d\varepsilon'\ \delta(\varepsilon - \varepsilon_j)\ \delta(\varepsilon - \varepsilon' - \varepsilon_i)$ where $A_{ij}$ corresponds to a bi-linear in the absorption parameters $a_{ij}$ (which are assumed to be $\sim 0.03\mu$); the delta functions correspond to spectral functions obtained from the Matsubara propagators of the Green's functions $g_{j\sigma}(\tau) = -\langle T\{a_{j\sigma}(\tau) a^\dagger_{j\sigma}(0)\}\rangle$ ( T is the time-ordering operator which arranges other operators from right to left in the ascending order of imaginary time $\tau$) using the formula $(x \pm i\ 0^+)^{-1} = [P(x^{-1}) \pm (1/i)\ \pi\ \delta(x)]$, where P represents a Cauchy's principal value. The quantity $\alpha_{ISBT}(\varepsilon)$ thus involves a bunch of delta peaks which may be replaced by the well-known result $\delta(x) = \lim_{\Gamma \to 0} (1/\pi)\ (\Gamma/(x^2 + \Gamma^2))$ to obtain a graphical representation. In Fig.2 we have also shown a plot of the absorption $\alpha_{ISBT}(\varepsilon)$ – vs –$\varepsilon$ at T= 300K. A fairly strong intersub-band absorption peak could be seen spread over the energy range 0.1-0.5 eV where $(\varepsilon_1 - \varepsilon_0)$ and $(\varepsilon_2 - \varepsilon_1)$ transitions do not contribute. Such transitions are normally forbidden by parity[22]. The corresponding emission peak (not shown) is roughly hundred times smaller at room temperature. We find that within a single-electron description, (even)in the absence of non-parabolicity and disorder, the inter sub-band absorption peak is not so narrow. This is due to the close positioning of the allowed transitions which leads to merging of the peaks into a single one.

It was first proposed by Barnham and Duggan [23] that the super-lattice-integrated solar cells have the potential to exceed the efficiency limits of the conventional single band-gap solar cells due to improved optical response of the cell in the energy region below the absorption edge of the well material. Later Anderson [24] and other workers have developed theoretical framework at semi-phenomenological level for ideal MQW solar cells for various material systems, such as AlGaAs/GaAs [25,26], InGaAs/ GaAs [27,28,29,30],etc.. Recently Aberhard and Morf [5] have applied non-eqilibrium Green's function approach for this problem in a model where two-band dispersion is captured approximately by placing an s-type orbital on the cation (Ga) and a $p_z$-type orbital on the anion (As). As can be seen from outline above that our effort is directed towards the investigation of AlGaAs/GaAs system, treating the absorption and emission together, with due emphasis on sub-band absorption. In order to examine its photovoltaic application, we consider a thick-emitter-thin-base variant of a p-i-n solar cell described in Ref.[31] by Aroutiounian et al.(APKT). We assume here n- and i-regions to be comparatively thinner compared to the p-region. In what follows we give a short discussion with result presentation. A practical cell of this type (hereinafter referred to as the 'test cell') based on III-V semiconductor materials, such as p-GaAs and n-GaAs, suggestively consists of n-and i-regions below p-GaAs layers with $z$ as the growth direction i.e., the normal to the layer plane. The quantum wells are located in the i-region. We, however, consider a p-i-n structure without the quantum wells (hereinafter referred to as the base cell) first. For this cell, the excess electron concentration $\Delta n(z,\lambda) = (n(z,\lambda) - n_0)$ ($n_0$ is the equilibrium concentration of electrons) in the p-type layer corresponding to a wavelength $\lambda$ in the incident photon flux satisfy the equation $D_n (d^2/dz^2)\Delta n = \{(\Delta n/\tau_n) - G(\lambda,z)\}$ where $D_n \sim 0.01$ m$^2$ s$^{-1}$ is the diffusion coefficient, G is the z-dependent photo-excitation rate ($G(\lambda,z) = G(\lambda,0)\exp(-\alpha_{bulk}(\lambda)z)$) takes into account the bulk absorption, through the coefficient $\alpha_{bulk}(\lambda)$, at any arbitrary point z and for a wavelength $\lambda$ in the incident photon flux), and $\tau_n$ (~0.1 ns) is the life-time of electrons. The z=0 plane is defined at the illuminated surface and the $z = z_1$ (= 1500 nm) plane is defined at the end of the base region. Furthermore, the electronic density starts off with a value $\delta n$ (=$\zeta N_D$ ($\zeta \ll 1$) ~ $10^{20}$ m$^{-3}$) at z=0, decreasing somewhat in the depletion layer, eventually attains a higher value approximately equal to the donor density $N_D$ (~ $5 \times 10^{21}$ m$^{-3}$) $\approx n_0$ in the n-region at z= $z_1$. The hole density, on the other hand, decreases sharply from $N_A$ (~$3 \times 10^{21}$ m$^{-3}$) and eventually attains a value ($n_i^2/N_D$) at z=$z_1$ where $n_i$ (~ $10^{14}$ m$^{-3}$) is the intrinsic concentration. Hence, the boundary conditions may be written as $(-eD_n)(d/dz)\Delta n\big|_{z=0} = J_n(\lambda,0)$, and $n(z_1) = \delta n \exp(qV/kT)$. Here $J_n(\lambda,0)$ is the electron current density corresponding to wavelength $\lambda$ due to incident photon flux creating free carriers at z=0. The quantity V is the forward voltage due to photo-excitation. At the assumed low doping density $N_D$, the space charge regions occupy a good portion of the device. We have solved the differential equation above and used the boundary condition $-eD_n (d/dz) \Delta n |_{z=0} = J_n(\lambda,0)$ to obtain $\Delta n$. We find $\Delta n(z,\lambda) = A(\lambda) \exp(zL_n^{-1}) + B(\lambda) \exp(-z L_n^{-1}) + C(\lambda)\exp(-\alpha z)$ where $L_n^2 = D_n \tau_n$,

$$A(\lambda) = (1/2)[\Delta n(0,\lambda) - (\tau_n/D_n)^{1/2} \{J_n(\lambda,0) + (G(\lambda,0)/(\alpha + L_n^{-1}))\}],$$

$$B(\lambda) = (1/2)[\Delta n(0,\lambda) + (\tau_n/D_n)^{1/2} \{J_n(\lambda,0) + (G(\lambda,0)/(\alpha - L_n^{-1}))\}],$$

and $C(\lambda) = -[G(\lambda,0)/D_n (\alpha^2 - L_n^{-2})]$. Here $\alpha$ stands for the absorption coefficient $\alpha_{bulk}(\lambda)$. Upon using the other boundary condition we obtain the current density $J = {}_0\int^{\lambda_m} d\lambda [-e\lambda_m^{-1} D_n (d/dz) \Delta n |_{z=z1}]$ in the standard form[31]: $J = [J_{sat} (\exp(qV/\nu kT) - 1) - J_{ph}]$ where $J_{sat}$ and $J_{ph}$, respectively, are the saturation current and the current produced due to the optical generation of carriers (also called the short-circuit current $J_{sc}$); $\nu$ is the ideality factor ~ 1. Here the saturation current $J_{sat}$ is formed by the minority carriers generated at the depletion layer edges and in the interior of i- region due to thermal excitation. The explicit expression of $J_{sat}$ is given in Ref.[31]. The upper limit $\lambda_m$ in the wavelength integration corresponds to the bulk band gap(~ 1.43 eV) of the material. As regards the test cell, since the absorption by the wells occurs at long wavelength region, the straightforward implication of this at the solar-cell-level is the enhanced utilization of the solar spectrum for the photo-carrier generation. In order to see how this would actually affect the carrier generation rate(CGR) in the of the i-region, with the aid of the absorption coefficient $\alpha_{ISBT}(\varepsilon)$ and Eq.(7) of Ref.[31], we calculate the photo-generated

12current contribution $J_{ph}$ from the 10 quantum wells located in the i-region which we assume to be of thickness $z_i$(~300 nm). The CGR for the wells, as in Ref.[31], is $G_{well}(\lambda,z) = -eF(\lambda)(1-R(\lambda))\alpha_{ISBT}(\lambda)\exp[-\alpha_{ISBT}(\lambda)z]$ where $F(\lambda)$ is the incident light flux and $R(\lambda)$ is the surface reflection coefficient. We have retained the same symbols (and numerical values) of Ref.[31] as far as possible. The photo-current collected from the entire i-region for a wavelength $\lambda$ of the incident light flux is $j_{well}(\lambda) = \int_0^1 G_{well}(\lambda,(z/z_i))\, d(z/z_i) = -eF(\lambda)(1-R(\lambda))\alpha_{ISBT}(\lambda)\{1-\exp[-\alpha_{ISBT}(\lambda)z_i]\}$. As will be seen below, the δ-peaks in $\alpha_{ISBT}(\lambda)$ yield a perceptible increase in the total photo current contribution $J_{ph}$ due to the additional term ($\int j_{well}(\lambda)\, d\lambda$).

It may be mentioned that, in their minority carrier transport equation based approach for a p-i-n cell where all the three regions are of comparable thickness, APKT were able to show that the inclusion of the quantum dots(QDs) in the intrinsic region does indeed enhance short-circuit current without the significant losses in the open circuit voltage and results in significantly improved cell efficiency. Their InAs/GaAs p-i-n solar cell, with 3 μm width of i-layer and the ideality factor v = 1.2, could generate the photo current density $J_{sc}$ = 45.17 mA cm$^{-2}$ and open circuit voltage $V_{oc}$ =0.746 V. Without the QDs, the same structure has $J_{sc}$ = 35.1 mA cm$^{-2}$ and $V_{oc}$ = 0.753 V. Accordingly, in the first case η = 25% while without QDs it is equal to 19.5%. As regards our results, we have shown J–V curves in Fig.3 without and with quantum wells at AM1 illumination with 1000W/m$^2$ intensity. We find that, due to the inclusion of absorption corresponding to the long wavelength region, there is increase in the short-circuit current and other photovoltaic performance related parameters: For the base cell, the short-circuit current $J_{sc}$= 184.3 A-m$^{-2}$, the open circuit voltage $V_{oc}$=0.8157V, the fill-factor (ff)=0.6086, and the power conversion efficiency ($\eta$)= 12.89 % at temperature 300 K. The outcomes of the similar calculation at the same temperature, for the test cell, are $J_{sc}$= 215.4 A-m$^{-2}$, $V_{oc}$=0.8354V, ff =0.7323,and $\eta$ = 15.48 %. Therefore, starting with a microscopic basis which we perceive as the strong point of this communication, we have been able to show that the quantum well involvement does give rise to a moderate increase in efficiency.

In summary, the investigations reported in this work may help one to understand the quantum well sub-band absorption leading to the additional carrier generation in III-V semiconductor based solar cells from a microscopic view-point, as we have used a multi-sub-band model (in one-electron picture) together with the Fermi Golden Rule approximation to describe the optical processes. A logical extension of this work is a rigorous analysis of the problem, which includes the correlation effect given by the term ($2U a_{i\uparrow}^{\dagger} a_{i\uparrow} a_{i\downarrow}^{\dagger} a_{i\downarrow}$), at a level beyond the mean-field (MF) approximation as the MF Hamiltonian is quadratic in fields and therefore exactly diagonalizable. Consequently, the corresponding eigen states are stationary states with infinite lifetime which precludes any possibility of finding whether the correlation effect is a contributory factor towards the absorption peak broadening. We hope to communicate in future the investigation carried out by us in this direction.

**16**

**Figure Captions**

Figure 1. A plot to show the variation of the conductance G with the increase in the ratio of Fermi energy and kT at a given magnetic field.

Figure 2. A plot of optical absorption coefficient of GaAs (bulk) and GaAs/AlGaAs quantum well.

Figure 3. J-V characteristics- the upper curve corresponds to a p-i-n cell without quantum wells (QWs) and the lower curve to one with QWs.

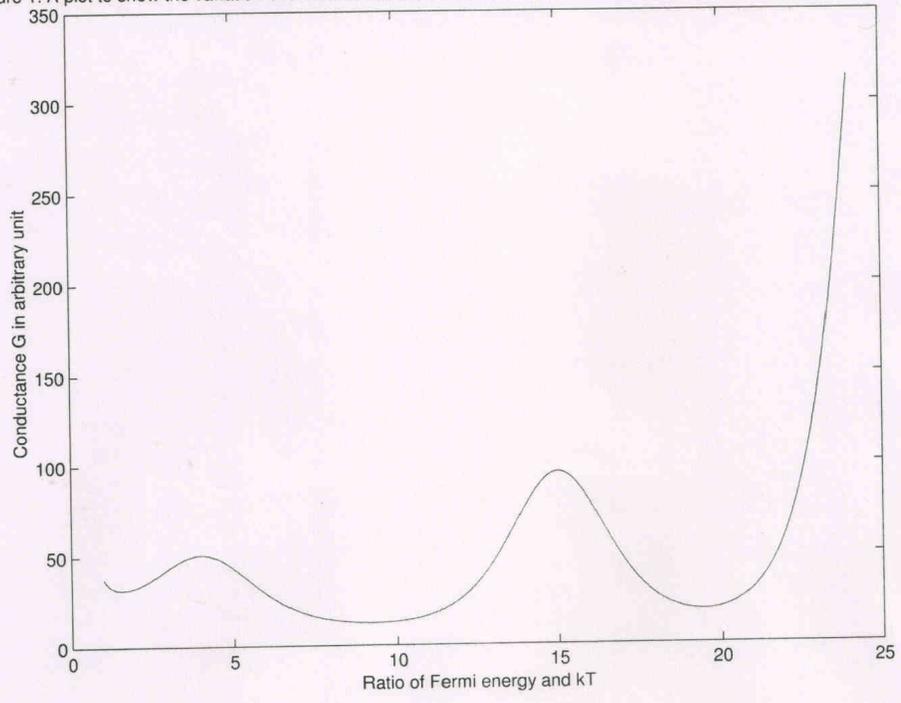

Figure 1: A plot to show the variation of the conductance G with the increase in the ratio of Fermi energy and kT at a given B.

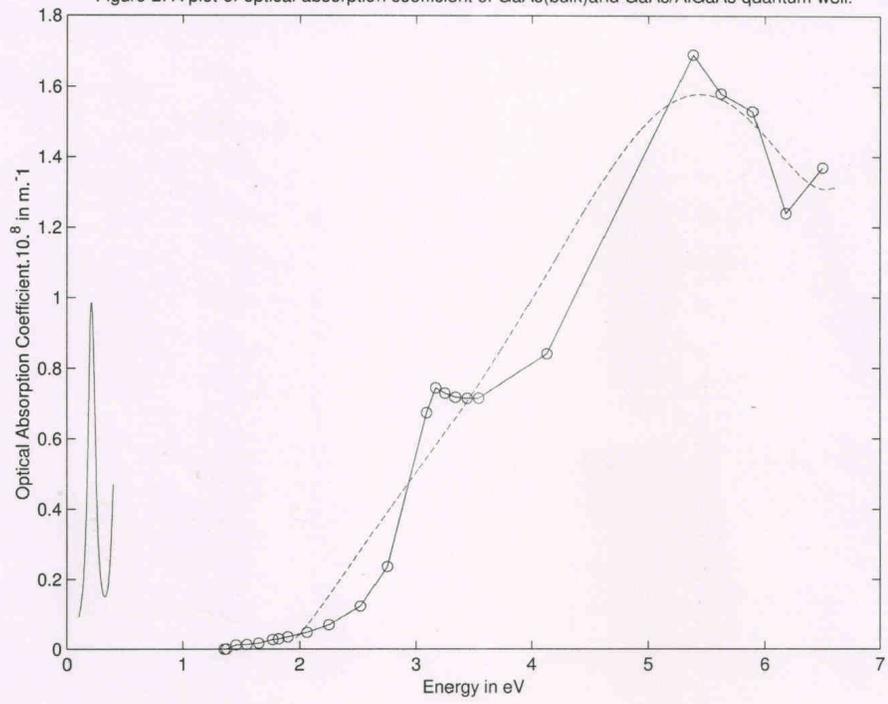
Figure 2: A plot of optical absorption coefficient of GaAs(bulk) and GaAs/AlGaAs quantum well.

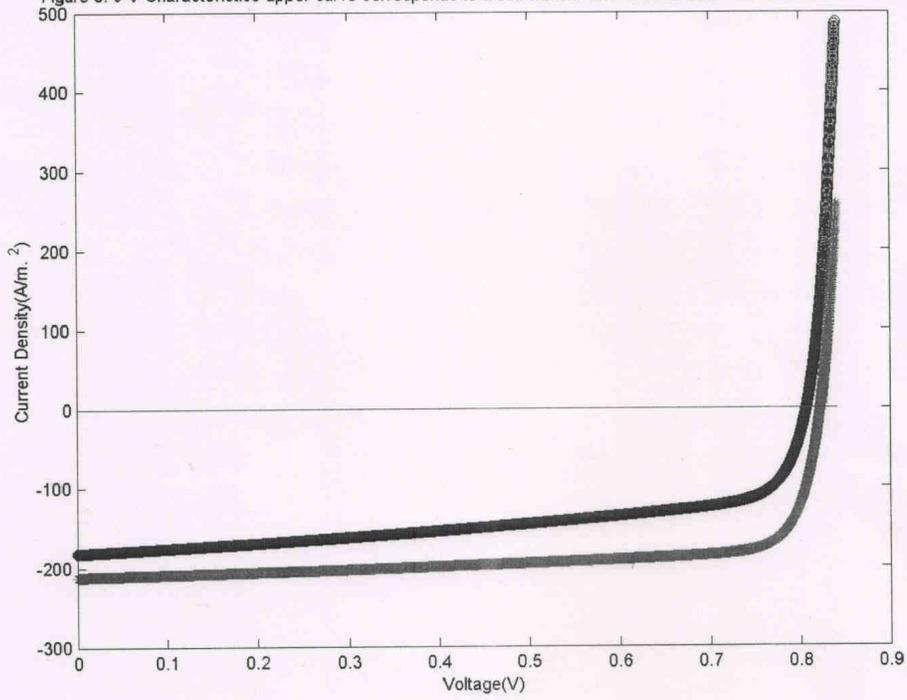

Figure 3: J-V Characteristics-upper curve corresponds to a cell without QW and the lower curve to one with QW.